\documentclass[%
 reprint,
 amsmath,amssymb,
 aps,
floatfix,
showkeys,
showpacs
]{revtex4-1}

\usepackage{graphicx}% Include figure files
\usepackage{dcolumn}% Align table columns on decimal point
\usepackage{bm}% bold math
\usepackage[section]{placeins}
\usepackage{makecell}
\usepackage[caption=false]{subfig}
\begin{document}

\preprint{APS/123-QED}

\title{Optimizing the $^{8}$Li yield for the IsoDAR Neutrino Experiment }
%\thanks{A footnote to the article title}

\author{Adriana Bungau}
%\email{abungau@mit.edu}
\author{Jose Alonso}
\author{Larry Bartoszek}
\author{Janet Conrad}
 \affiliation{Massachusetts Institute of Technology, Cambridge, 
Massachusetts, MA 02139, United States of America}

\author{Michael Shaevitz}
\affiliation{Columbia University, New York, Irvington, NY 10533, United
States of America}
 
\author{ Joshua Spitz}
\affiliation{University of Michigan, Ann Arbor, MI 48109, United
States of America}
 
\date{\today}% It is always \today, today,

\begin{abstract}

The focus of this paper is on optimizing the electron-antineutrino 
source for the IsoDAR (Isotope Decay at Rest) experimental program. 
IsoDAR will perform sensitive short-baseline neutrino oscillation and 
electroweak measurements, among other Beyond Standard Model searches, 
in combination with KamLAND and/or other suitable detectors. IsoDAR 
will rely on the high-$Q$ $\beta^-$ decay of the $^{8}$Li isotope for 
producing electron-antineutrinos, created mainly via neutron capture in 
an isotopically enriched $^{7}$Li sleeve surrounding the Be target. In 
particular, this paper examines the performance, defined in terms of 
absolute $^{8}$Li (or, equivalently, electron-antineutrino) production 
rate, of various candidate sleeve materials, including a lithium-fluoride, 
beryllium-fluoride mixture (``FLiBe'') sleeve and a homogeneous 
mixture of lithium and beryllium (``Li-Be''). These studies show that the 
$^{8}$Li yield can be increased substantially by employing a Li-Be sleeve 
and therefore motivate significant changes to the nominal IsoDAR design.

\end{abstract}

%\pacs{pacs_no}


\maketitle

\section{Introduction}

\vspace{0.5 cm}

\noindent
The three neutrino paradigm works extraordinarily well in describing 
mixing between the established flavor states. However, a number of 
fundamental questions remain. For example, a number of anomalous 
results, including those from LSND~\cite{LSND}, MiniBooNE~\cite{MiniBooNE} 
and short-baseline reactor experiments~\cite{sb-reactor}, may be 
interpreted in the context of mixing involving one or more sterile 
neutrinos, a possibility which has excited the physics community. 
These inconsistencies may be explained by a (3+N) sterile neutrino 
model in which there are three light neutrino mass states and N 
massive sterile neutrino mass states. 
 
\noindent
The IsoDAR experimental program is the first stage of the 
DAE$\delta$ALUS experiment~\cite{isodar_prl,conrad_and_shaevitz} 
and is being developed to investigate the anomalies possibly 
indicative of neutrino oscillations at high-$\Delta m^2$. IsoDAR 
represents an important technological step, in terms of producing 
high-power cyclotrons for a number of physics and non-physics 
applications, and physics reach, in the context of resolving the 
sterile neutrino anomaly question and providing a number of 
electroweak and exotic search measurements~\cite{electroweak}. 
The IsoDAR experiment is based on a high intensity $\bar{\nu}_{e}$ 
source, originating from the $\beta^-$ decay of $^{8}$Li, 
coupled to a massive scintillator-based detector (e.g. KamLAND). 
The $\bar{\nu_{e}}$ can interact in the detector via the inverse beta 
decay (IBD) process: $\bar{\nu}_{e}$ + $p$ $\rightarrow$ $e^{+}$ + $n$.
IsoDAR will be sensitive to mapping the $\bar{\nu}_{e}$ ($<$$E>$ = 6.4 
MeV) disappearance oscillation wave and will study the possibility of 
high-$\Delta m^2$ mixing, with the unique ability to distinguish 
between models with one or two sterile neutrino flavors.

\subsection{The IsoDAR experiment} 

\noindent
The IsoDAR neutrino source is created with high current (5~mA), 
60~MeV/amu H$^{+}_{2}$, which produces the neutrons required for 
the production of $^8$Li, and resulting $\bar{\nu_{e}}$. An ion source 
is used to produce the H$^{+}_{2}$, which are then accelerated via a 
low energy beam transport system and a cyclotron, before being extracted 
to the target. The IsoDAR experimental program will use the high power 
DAE$\delta$ALUS injector cyclotron to produce this intense isotope 
decay-at-rest antineutrino source. The neutrons produced from the 
H$^{+}_{2}$ interactions with the heavy-water-cooled $^{9}$Be primary 
target create $\bar{\nu}_{e}$ from capture in an isotopically enriched 
($99.99\%$) $^{7}$Li target sleeve. This degree of enrichment
is needed to avoid production of tritium by neutron capture on $^{6}$Li, as
this cross section is several orders of magnitude larger than that of
neutron capture on $^{7}$Li. 

\noindent
Inverse beta decay interactions (IBD; $\bar{\nu}_{e}p \rightarrow e^+ n$) from 
this uniquely efficient target design are then collected by a $\sim 1$~kiloton 
scintillator-based detector at a distance of about $\sim 16$ m. The 
existing KamLAND detector would be suitable, making this proposed 
experiment also cost-effective compared to other approaches. Five years 
of running will provide nearly a million IBD events, suitable to measure the 
coupling parameter $\sin^{2} \Theta_{w}$ to several percent and to 
definitely determine if there are 0, 1, or 2 light sterile neutrinos.

\section{The Conceptual Design of the Target System}
  
\subsection{The current target design}

\begin{figure}[!h] 
\centering
\includegraphics[width=0.45\textwidth]{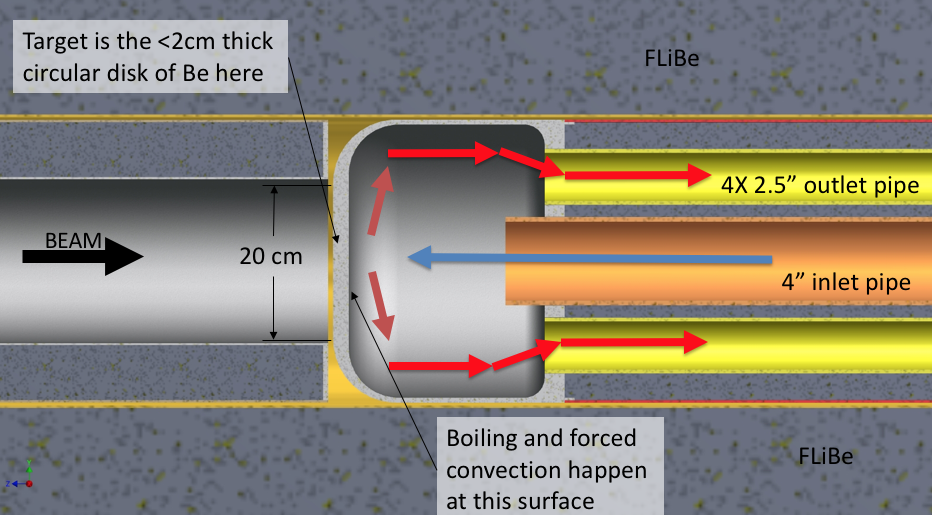}
\caption{Detail of the target region showing the direction of cooling
  water flow. }
\label{fig:1}
\end{figure}

\noindent
The current target design~\cite{CDR} consists of a hollow Be 
cylinder leading to a circular Be disk (Fig.~\ref{fig:1}). The peak 
power density in the target is decreased by a wobbler magnet 
placed in front of the target-shielding system which spreads 
the beam onto its 20~cm diameter target face. The cooling 
system is at the back end of the target, with high pressure heavy 
water introduced along the central axis, and providing dissipation 
of the 600~kW beam power deposited on the target. The water is
led away by pipes located radially around the inlet pipe. 

\begin{figure}[!h] 
\centering
\includegraphics[width=0.49\textwidth]{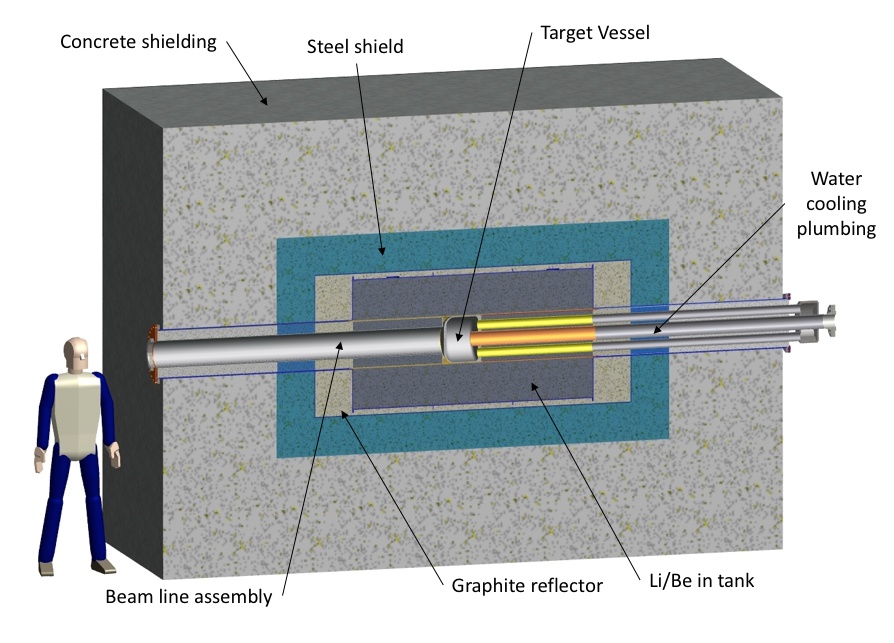}
\caption{Cross section of the target and shielding
  assembly.}
\label{fig:2}
\end{figure}

\noindent
The target is surrounded by a 1~m outer diameter and 1.9~m length 
sleeve (Fig.~\ref{fig:2}). In the current design \cite{CDR}, the sleeve 
material is a mixture of lithium-fluoride and beryllium-floride (FLiBe), 
an eutectic mixture with a melting point of $459^{\circ}$C and a density 
of $1.94$~g/cm$^{3}$. The low atomic weight of Li, Be, and to a lesser 
extent F, makes FLiBe an effective neutron moderator. 

\vspace{0.2 cm}

\noindent
FLiBe is commonly used as a coolant in molten-salt reactors. The  $^{7}$Li 
isotopic purity assumed is $99.995\%$ although simulation studies have 
shown that $99.99\%$ enrichment can meet the physics requirements 
of the experiment in terms of $\bar{\nu}_e$ production~\cite{Bungau}. 
A 5~cm thick graphite neutron reflector surrounds the sleeve. The 
reflector is meant to redirect the neutrons back into the sleeve where 
they can be further used for $^{8}$Li production. The entire target and 
sleeve system is enclosed in shielding, with iron inner layers and boron 
rich concrete outer layers.

\subsection{The modelling of the target system} 

\noindent
The target system is modelled using the Monte Carlo code 
Geant4~\cite{geant4} where the hadronic interactions of incident 
and secondary particles with target materials, and energy ranges 
are defined in the physics package \textit{particle\_{hp}}. This 
package uses the \textit{ENDF/B-VII.1}~\cite{ENDF} and 
\textit{TENDL-2014}~\cite{TENDL} libraries for cross sections of 
primary and secondary particle interactions. The \textit{ENDF/B-VII.1} 
library uses experimental data for projectile energies up to 150 MeV. 
These data are essentially nuclear reaction cross sections together 
with the distributions in energy and angle of secondary reaction 
products. The \textit{ENDF/B-VII.1} database for proton projectiles 
contains data only for 48 isotopes (including Be). The \textit{TENDL-2014} 
library uses some experimental data and \textit{TALYS}~\cite{TALYS} 
calculations for projectile energies up to 200 MeV. The database can 
be applied to all target materials but the best results are obtained for 
targets with atomic number in the range 12-339. The \textit{TENDL-2014} 
database contains information for all isotopes.

\subsection{$^{8}$Li isotope production} 

\begin{figure}[ht!]
  \begin{center}
          \subfloat[ 2 cm target thickness \label{fig:3a} ]
          %  {\includegraphics[width=73mm, height=47mm]{/isotopes_per_proton.png}}\\
            {\includegraphics[width=93mm, height=47mm]{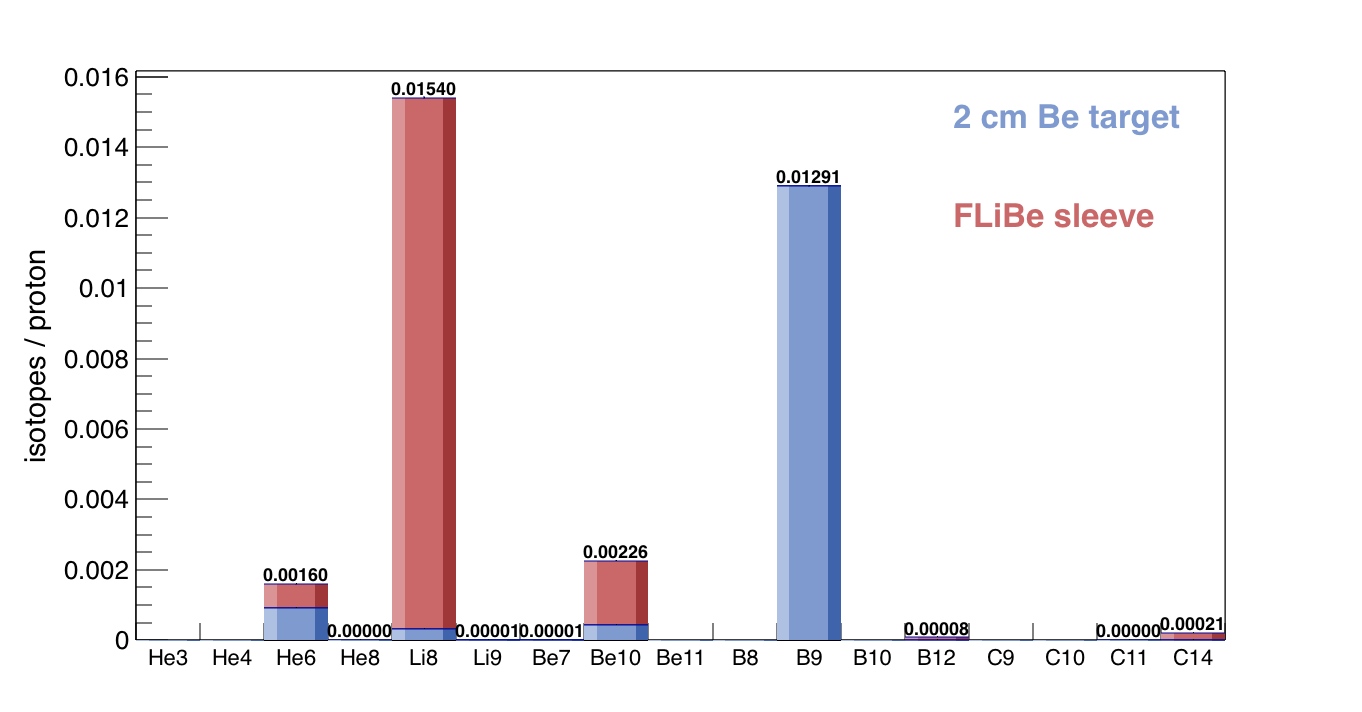}}\\
         \subfloat[ 1.7 cm target thickness \label{fig:3b} ]
        %    {\label{fig:3b}\includegraphics[width=83mm, height=48mm]{/FLiBe_17mm.png}}
            {\includegraphics[width=93mm, height=48mm]{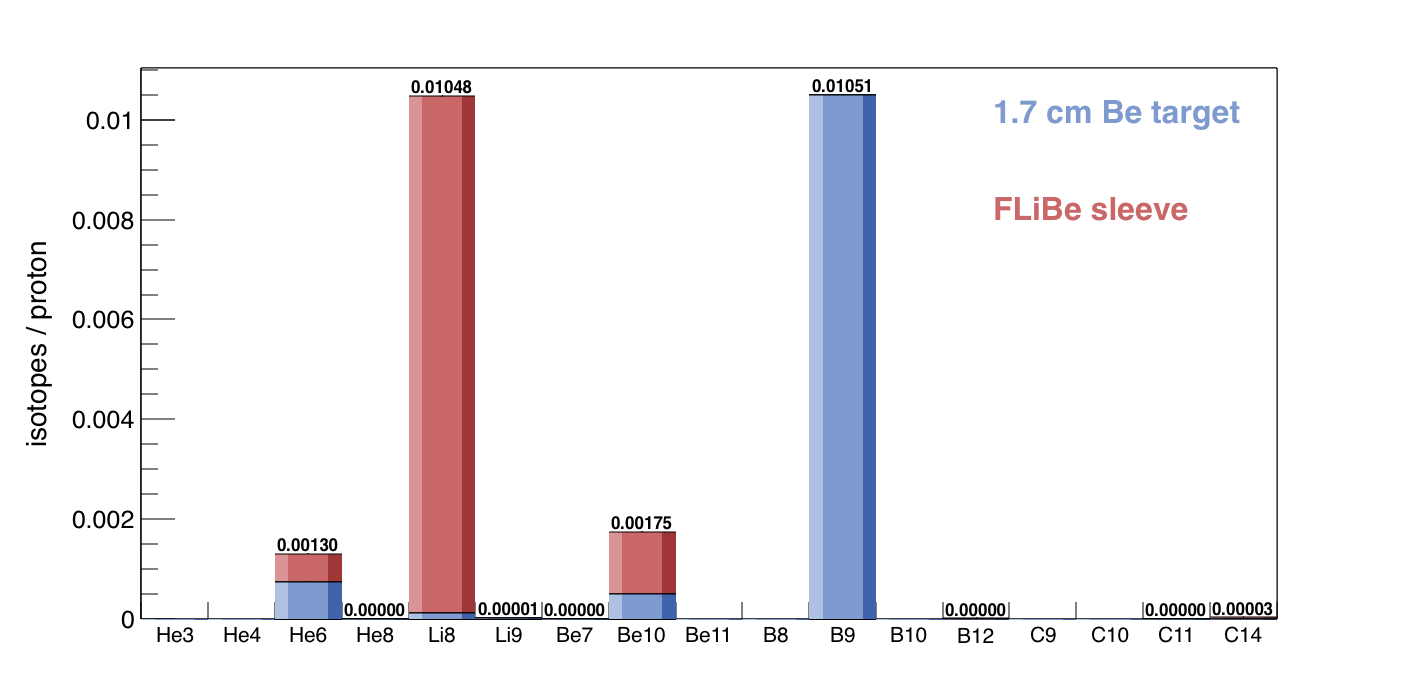}}
\end{center}
  \caption{The $^{8}$Li production in the target and sleeve, among
    other isotopes, for 2~cm target thickness 
  and 1.7~cm target thickness.}
  \label{fig:3}
\end{figure}

\noindent
The isotope production for incident 60~MeV protons on the target 
and sleeve configuration described is shown in Fig.~\ref{fig:3}. 
For a target thickness of 2 cm, the overall $^{8}$Li production is 
0.015~$^{8}$Li per incident proton, with most of the $^{8}$Li produced 
primarily inside the FLiBe sleeve. The Bragg peak of 60~MeV protons 
in Be is at 2.04~cm and the 600~kW of beam power deposited on 
the target can seriously affect the sustainability of the target 
performance in consideration of a 5~year physics run. This issue 
has been addressed by reducing the target thickness. The new target 
thickness of 1.7~cm is a compromise between maximizing neutron 
production and decreasing heat deposition in the Be. The $^{8}$Li 
isotope production drops to 0.010 $^{8}$Li per proton in this design, 
however. As the main contribution to $^{8}$Li production comes from 
the sleeve, studies of sleeve material and geometry optimization have 
been performed and the results are presented below.

\section{Optimization of the Sleeve}

\subsection{Neutron interactions inside the FLiBe sleeve}

\begin{figure*}[!htp] 
\centering
\includegraphics[width=0.9 \textwidth]{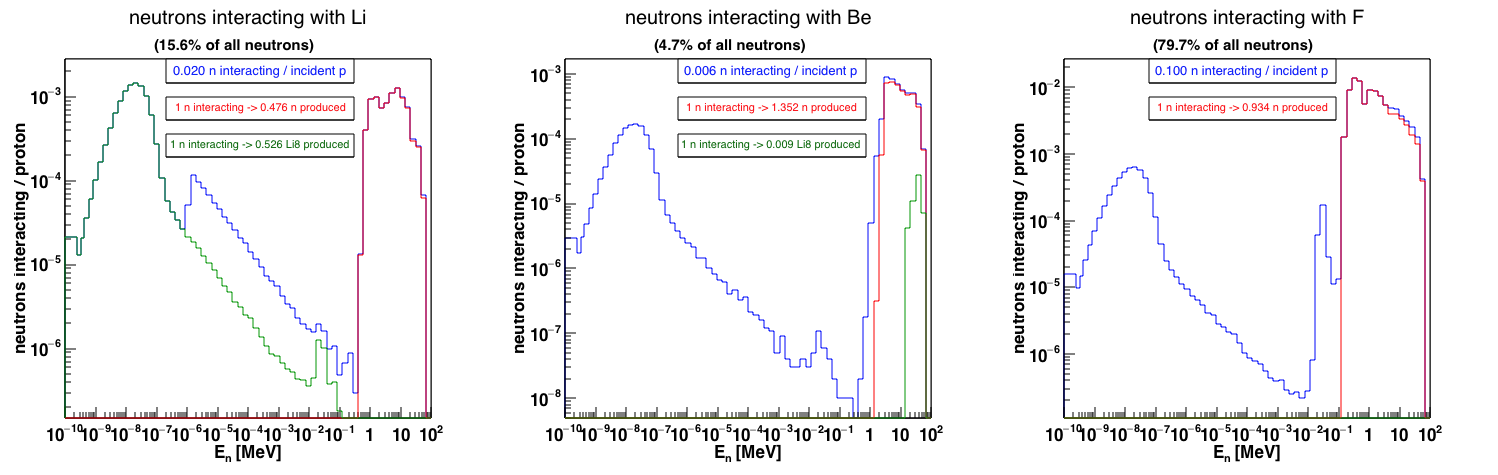}
\caption{The energy spectrum of all neutrons interacting inside the
  FLiBe sleeve is shown separately for neutron interactions on Li, Be, 
  and F (blue). Some of the neutrons interact to produce other 
  neutrons (red) and some interact to produce $^{8}$Li (green).}
\label{fig:4}
\end{figure*}

\noindent
The neutron interactions of interest inside the FLiBe sleeve are 
(1) inelastic processes in which other secondary particles are 
produced and (2) neutron capture processes which lead to isotope production. 
The $^{8}$Li is produced via direct production processes like neutron 
capture on $^{7}$Li (98.3$\%$ of the total $^{8}$Li production), or by 
neutron inelastic interactions with Be in which a neutron and a proton 
are produced (1.7$\%$ of the total $^{8}$Li production). The production 
can be increased via indirect processes in which secondary neutrons are 
produced with a neutron multiplication factor higher than 1. These secondary 
neutrons can interact at a later stage inside the sleeve and 
contribute to the direct $^{8}$Li production processes mentioned 
above. The neutrons interact with all three elements present inside the 
sleeve; the contribution of each element to the overall $^{8}$Li
production is shown in Fig.~\ref{fig:4}. The energy spectra of all the 
neutrons interacting either inelastically or by neutron 
capture on Li, Be and F are shown separately in blue.

\noindent
A large fraction of neutrons interact with F (80$\%$),  while a 
much smaller number of neutrons interact with Li ($\approx$15$\%$) 
and Be ($\approx$5$\%$). A fraction of these neutron interactions 
result in direct $^{8}$Li production, and the energy spectra of these 
neutrons which produce $^{8}$Li are shown in green in Fig.~\ref{fig:4}. 
The main production mechanism for these comes from the capture of 
$<$0.1 MeV neutrons on $^{7}$Li (left plot). For interactions on Be, 
the $^{8}$Li is produced in inelastic processes and the neutrons have 
energies above 10 MeV. Importantly, there are no processes in which $^{8}$Li is 
produced as a result of neutron interactions with F. 

\noindent
Another fraction of neutrons produced inside the sleeve come from neutrons 
interacting inelastically and producing other secondary neutrons. 
The spectra of the neutrons producing additional secondary neutrons are 
shown separately in red.  The only neutron interactions which have 
an average neutron multiplication factor greater than 1
are the inelastic interactions on Be. In the case of Be, for each
neutron used inside the sleeve by any process - either capture or
inelastic, there will be $<$1.35$>$ secondary neutrons produced.

\noindent
Regarding the neutron interactions on Li, there are 0.02 neutron interactions
(capture or inelastic) per incident proton on target. Out of these interactions, 
$\approx$53$\%$ are low-energy neutron capture processes on $^{7}$Li, and 
$\approx$47$\%$ are inelastic processes that produce additional neutrons. 
Regarding the neutron interactions on Be, there are 0.006 neutron interactions 
on Be per incident proton. Figure ~\ref{fig:4} shows that there is a significant number of low 
energy neutrons that are captured on Be which produce other products that 
are not relevant for IsoDAR. However, Be contributes to neutron multiplication as
there are $<$1.35$>$ neutrons produced for every neutron interaction as 
mentioned above. Only a small fraction of high energy neutrons produce 
$^{8}$Li in Be. F does not contribute at all to the direct $^{8}$Li 
production. The neutron production in F is $<$0.93$>$ neutrons per neutron 
interaction. Therefore, the number of neutrons inside the sleeve is slightly 
decreased following each interaction. The first blue peak corresponds to 
neutron capture on F, and the second peak corresponds to
inelastic neutron interactions.

\begin{figure*}[!htp] 
\centering
\includegraphics[width=0.9\textwidth]{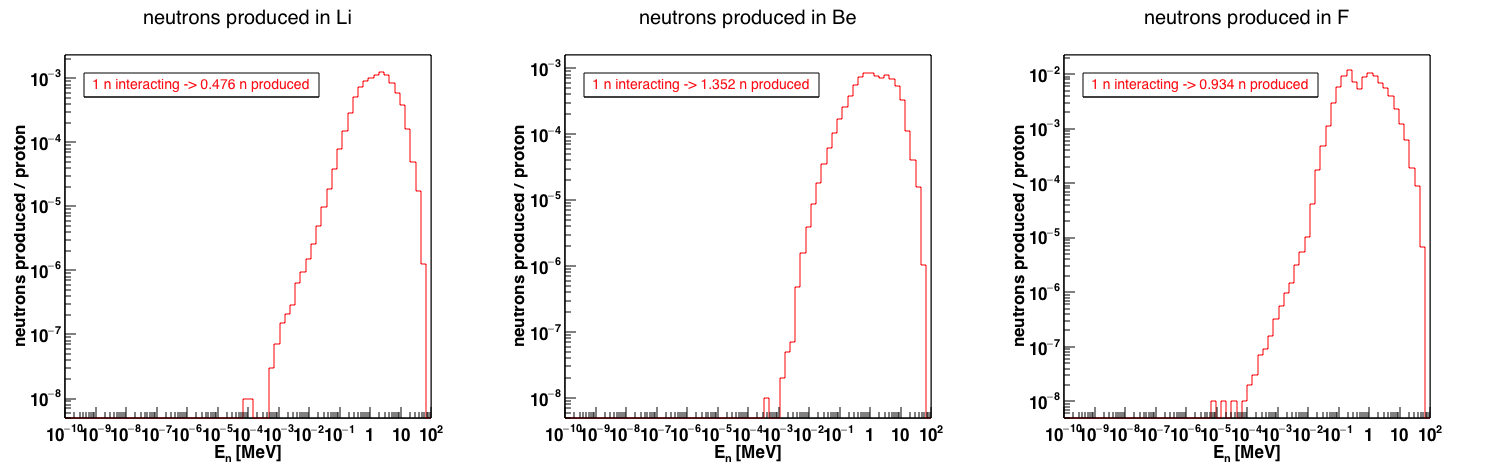}
\caption{The energy spectra of neutrons produced in interactions on
  Li, Be and F. }
\label{fig:5}
\end{figure*}

\noindent
Apart from the energy spectrum of the neutrons used 
inside the sleeve, a second factor of interest in this study is the
energy spectrum of the neutrons produced in inelastic
processes. These neutrons can be further used in neutron capture 
processes on $^{7}$Li that produce $^{8}$Li. Figure~\ref{fig:5} shows 
the energy spectra of the neutrons that are produced in the sleeve. 
They are similar to the energy spectra of the neutrons which interacted 
inelastically to produce them, however they show a small tail towards 
the low energy end.  

\subsection{Neutron interactions inside the Li-Be sleeve}  

\begin{figure*}[!htp] 
\centering
\includegraphics[width=1.\textwidth, height=0.23\textheight]{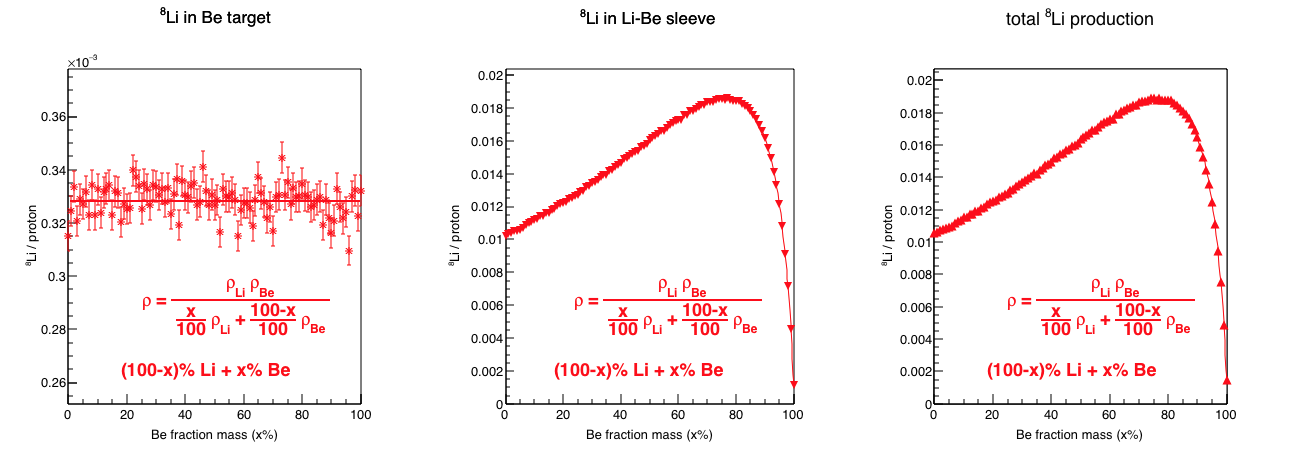}
\caption{The overall $^{8}$Li production in the target and sleeve, for
  various concentrations of Li and Be.}
\label{fig:6}
\end{figure*}

As the atomic ratio of Li:Be:F is 2:1:4, the presence of F drastically 
reduces the atomic number density of both Li and Be inside the sleeve, 
and therefore the number of neutron interactions with Li and Be that 
can produce $^{8}$Li. Since F does not contribute to direct 
$^{8}$Li production, and the production of secondary neutrons is 
decreased after each neutron-F interaction, the study of material 
optimization has been focused on a sleeve made only of Li and Be. In 
order to determine the optimum composition of such a Li-Be sleeve, a
parametric study for variable Be mass fraction (by weight) was
performed. The results are shown in Fig.~\ref{fig:6}. As can be seen, the
optimum design consists of a homogeneous mixture of enriched $^{7}$Li
and Be, with a Be mass fraction of 75 $\%$ (0.019~$^{8}$Li per incident proton). 

\noindent
The $^{8}$Li rates mentioned above are obtained for a homogeneous 
mixture of lithium and beryllium. However, in practice, a mixture close 
enough to homogeneity can be obtained by using small beryllium spheres 
immersed in lithium, and preliminary studies have shown that the $^{8}$Li 
production increases with decreasing radius. A parametrization study of 
$^{8}$Li yield versus radius was performed in order to find out how much 
the radius should be decreased and what is the correspondent value of 
$^{8}$Li production. 

\noindent
Due to the complexity of the code required to simulate a very large number 
of Be spheres stacked inside the Li sleeve cylinder, avoiding at the same time 
all the inner volumes (e.g. Be vessel and cooling pipes), instead of the actual 
sleeve, a fixed size box of lithium of (120x120x120) cm$^3$ was simulated 
to approximate the sleeve. The box was filled with beryllium spheres of various 
radii. The neutrons were generated in the centre of the box using the same 
energy spectrum as the neutrons entering the actual sleeve. 

\noindent
The aim of this study was to find out how small the Be spheres have to
be in order to approximate the Li-Be homogeneity, in order to ensure
that the predicted $^{8}$Li yields from Fig.~\ref{fig:6} can be achieved in practice when
using stacked Be spheres, rather than a homogeneous Li-Be mixture.

\begin{figure*}[!htp] 
\centering
\includegraphics[width=0.6\textwidth]{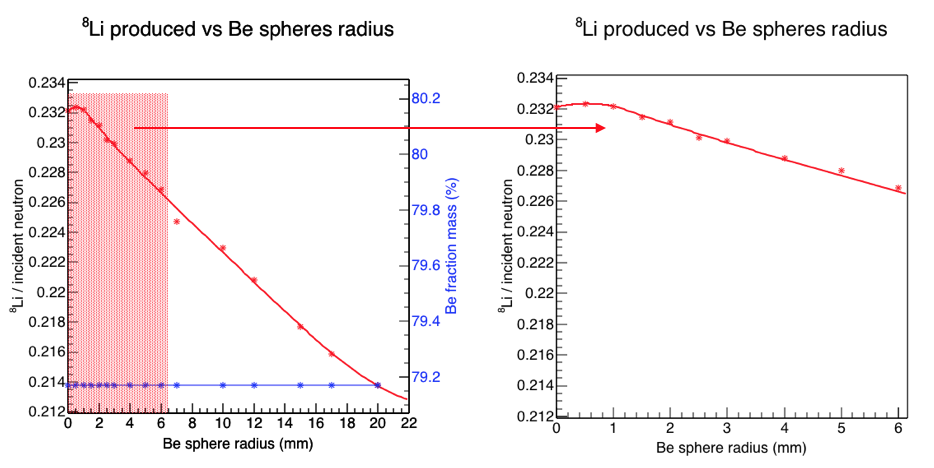}
\caption{The $^{8}$Li production in the Li-Be cube, for
  various radii of the Be spheres.}
\label{fig:7_radii}
\end{figure*}

\noindent
The results can be seen in fig.~\ref{fig:7_radii} and they show a
continuous steadily increase in the $^{8}$Li yield with decreasing sphere radius, 
all the way down to 1 mm radius. Below 1 mm, a plateau is reached and
no further increase in the $^{8}$Li yield can be observed. A 
homogeneous mixture of Li and Be inside the same (120x120x120)
cm$^3$ cube was also simulated, and it was found that the $^{8}$Li
yield was the same as for the 1 mm radius spheres design.

\noindent
The Beryllium fraction mass was calculated for each sphere radius and it
was found to be $\approx$79\% in all cases due to the level of
compactness of the spheres regardless of their radii, value which is
within the optimum range found for the homogeneous Li-Be mixture. 

\noindent
If the radius is decreased from 20 mm to 1 mm, this results in a ~10\% 
increase in the $^{8}$Li production. Therefore having a Li sleeve filled with 
Be spheres of $\sim$1 mm radius will result in an optimum configuration 
regarding the total $^{8}$Li production. 
  
\begin{figure*}[ht!]
  \begin{center}
            \subfloat[ neutrons interacting with Li and Be \label{fig:8a} ]
             { \includegraphics[width=93mm,
               height=50mm]{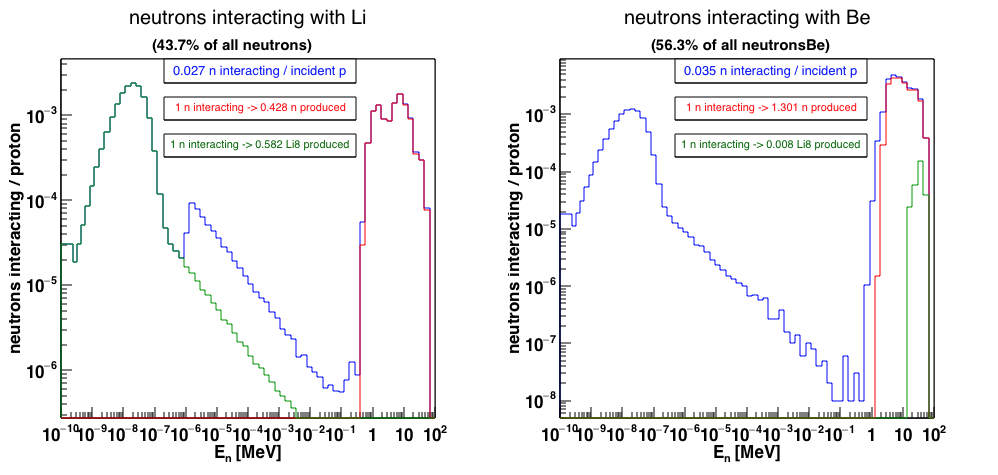}}
             \subfloat[ neutrons produced in Li-Be \label{fig:8b} ]
            {\includegraphics[width=84mm, height=49mm]{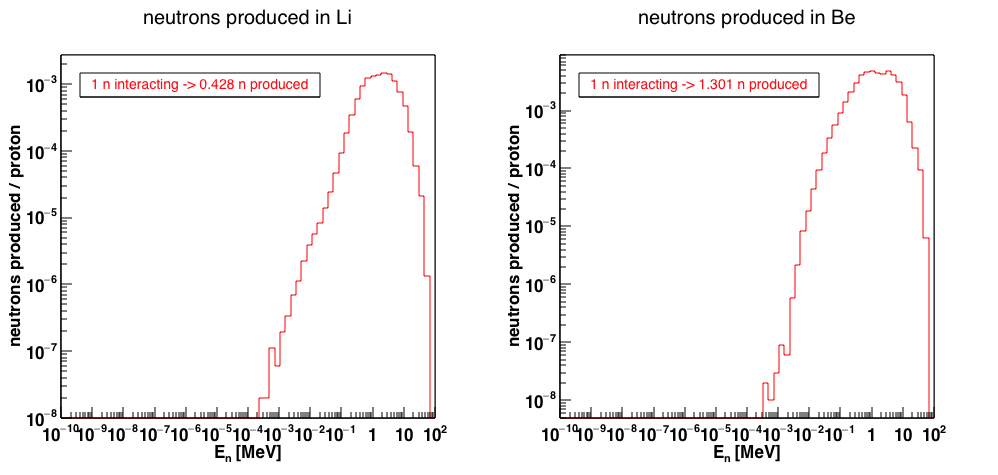}}
\end{center}
  \caption{The energy spectrum of all neutrons interacting
  in a Li-Be sleeve with a Be mass fraction of 75$\%$, is shown separately for neutron interactions on
  Li and Be (blue). A fraction of the neutrons interact producing other 
  neutrons that can be further used (red) and a fraction interact to
  produce $^{8}$Li (green). The energy spectra of neutrons produced in interactions on
  Li and Be are also shown.}
  \label{fig:8}
\end{figure*}

\noindent
In the Li-Be sleeve, the neutrons interact in the same way with   
Li and Be as they do with the Li and Be 
inside the FLiBe sleeve. However, what changes significantly 
is the number of neutrons interacting with Be per incident
proton, and this number has increased by a factor of six
(Fig.~\ref{fig:8a}). For the Li-Be sleeve there are 0.035~neutron
interactions with Be per incident proton, and the average neutron
multiplication factor in Be is 1.3~neutrons produced per neutron
interaction. Similarly, the energy spectra of the neutrons that are 
produced in the sleeve is shown in Fig.~\ref{fig:8b}
with a small tail towards the low energy end.  

\subsection{Change in the $^{8}$Li production for two sleeve designs}
 
\begin{figure}[ht!]
  \begin{center}
            \subfloat[ $^{8}$Li production in Li \label{fig:9a} ]
             {\includegraphics[width=46mm, height=44mm]{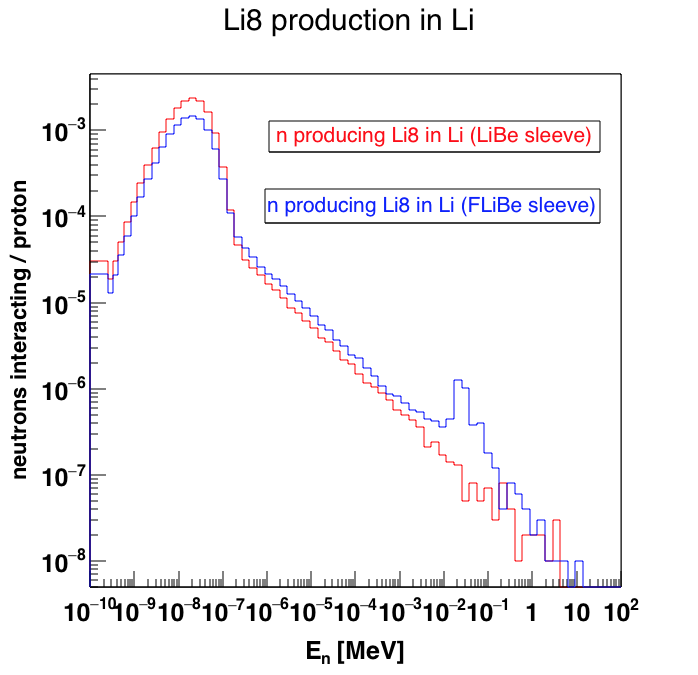}}
             \subfloat[$^{8}$Li production in Be\label{fig:9b}]
              {\includegraphics[width=46mm, height=44mm]{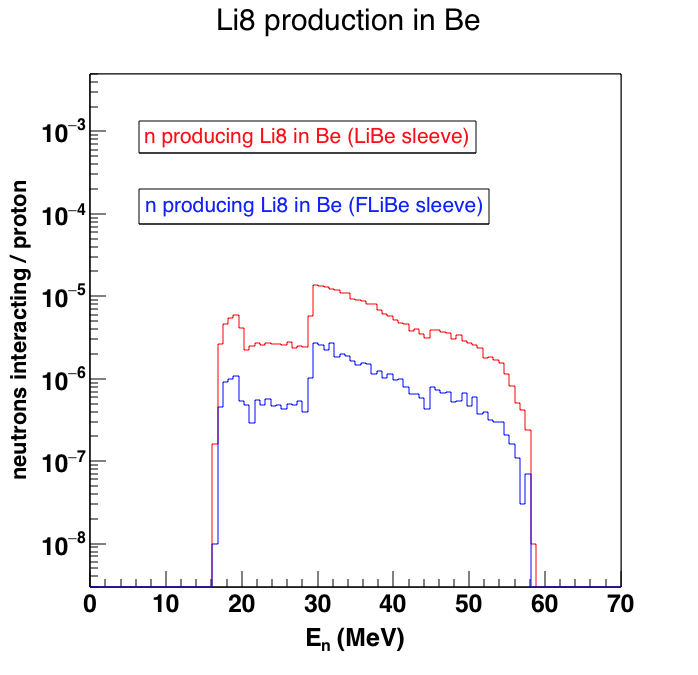}}
\end{center}
  \caption{Change in the $^{8}$Li production for the two sleeve materials. The $^{8}$Li production is shown separately for
  interactions on Li and Be.}
 \label{fig:9}
\end{figure}

Since the number of neutron interactions with the Be nuclei increases by a 
factor of 6 due to the increase in the atomic number density of 
Be inside the sleeve, the production rate of $^{8}$Li from neutron 
inelastic interactions with Be also increases by the same
factor. This can be seen in Fig.~\ref{fig:9b}. However, the number 
of neutron interactions with Li nuclei does not change significantly 
between the two target sleeves as seen in Fig.~\ref{fig:9a}. 
While in the case of Be interactions, the $^{8}$Li is produced in
inelastic processes at high neutron energies, in the case of 
neutron interactions on Li, the neutron capture process on 
$^{7}$Li produces $^{8}$Li, and the cross-section increases 
with decreasing neutron energy.

\begin{figure}[!ht] 
\centering
\includegraphics[width=0.5\textwidth]{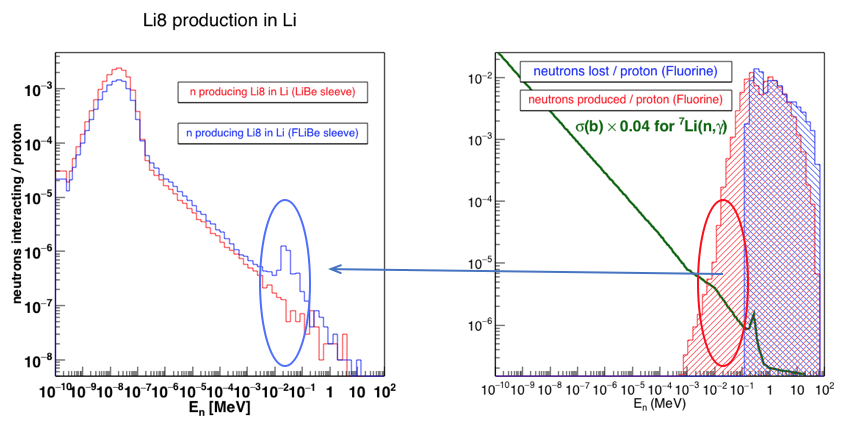}
\caption{ An analysis of the peak in $^{8}$Li
  production in the FLiBe sleeve.}
\label{fig:10}
\end{figure}

\noindent
There are two distinct differences in the $^{8}$Li production between the 
two sleeves. When F is present in the sleeve there is 
a high peak for neutron energies in the range 10 - 100 keV.  
This difference is explained in Fig.~\ref{fig:10} which shows the energy 
spectrum of the neutrons lost in inelastic interactions with F (blue), 
and the energy spectrum of neutrons produced in these interactions 
(red). It can be seen that, as a result of the inelastic interactions with F, 
there is a surplus of neutrons in the sleeve in the energy range 1 - 100 keV. 
Some of these neutrons are captured on $^{7}$Li, producing $^{8}$Li, 
as there is a finite cross-section (shown in green in Fig.~\ref{fig:10}) for 
this process at these neutron energies.

\noindent
Apart from this surplus of neutrons in the 1 - 100~keV energy range, there 
is also a slight decrease in the number of neutrons in the very low energy 
range (below 10$^{-7}$ MeV) due to the presence of F in the sleeve
(Fig.~\ref{fig:9a}). For these energies, some of the neutrons will 
be lost in neutron capture processes on F, so the number of 
interactions with $^{7}$Li will also decrease when F is present.

\subsection{Changes to the sleeve design} 

\begin{figure*}[hbp!]
  \begin{center}
                     {\label{fig:f01}\includegraphics[width=59mm, height=66mm]{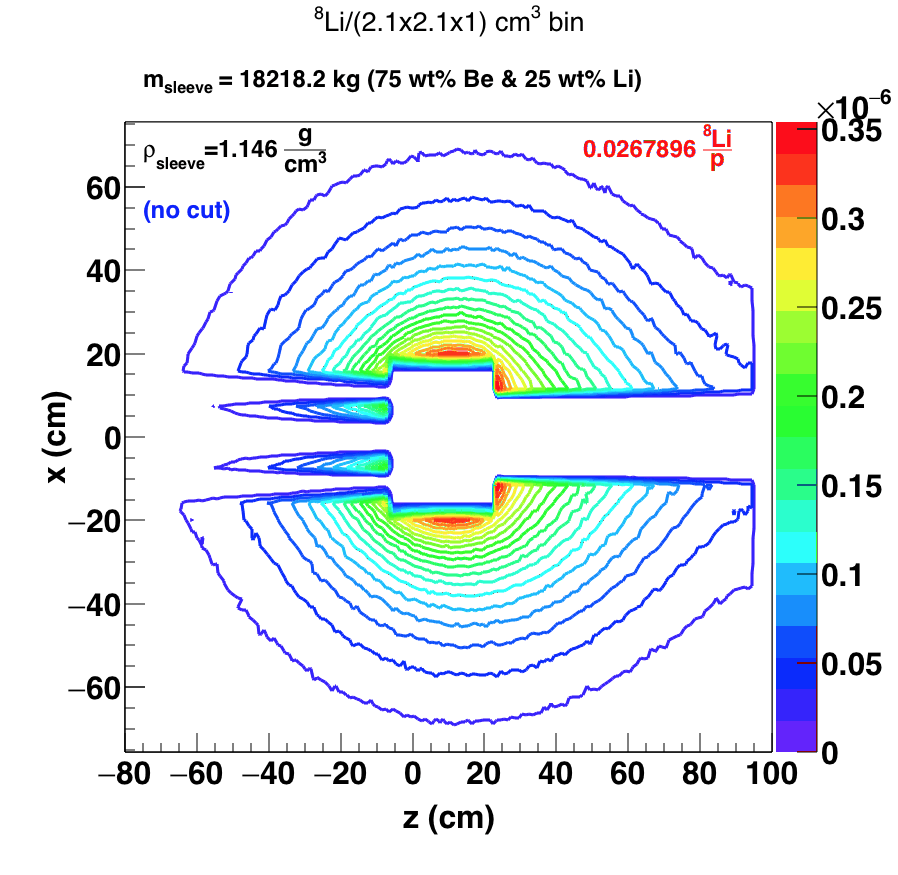}}
                     {\label{fig:f02}\includegraphics[width=59mm, height=67mm]{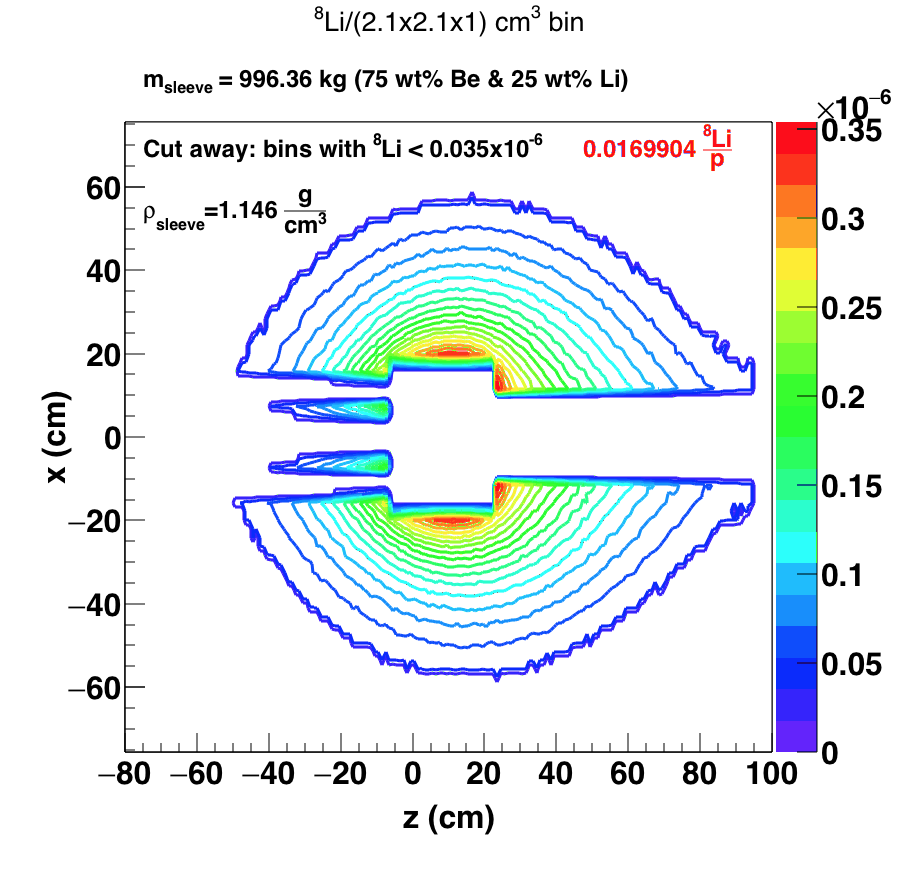}}
                     {\label{fig:f03}\includegraphics[width=59mm, height=67mm]{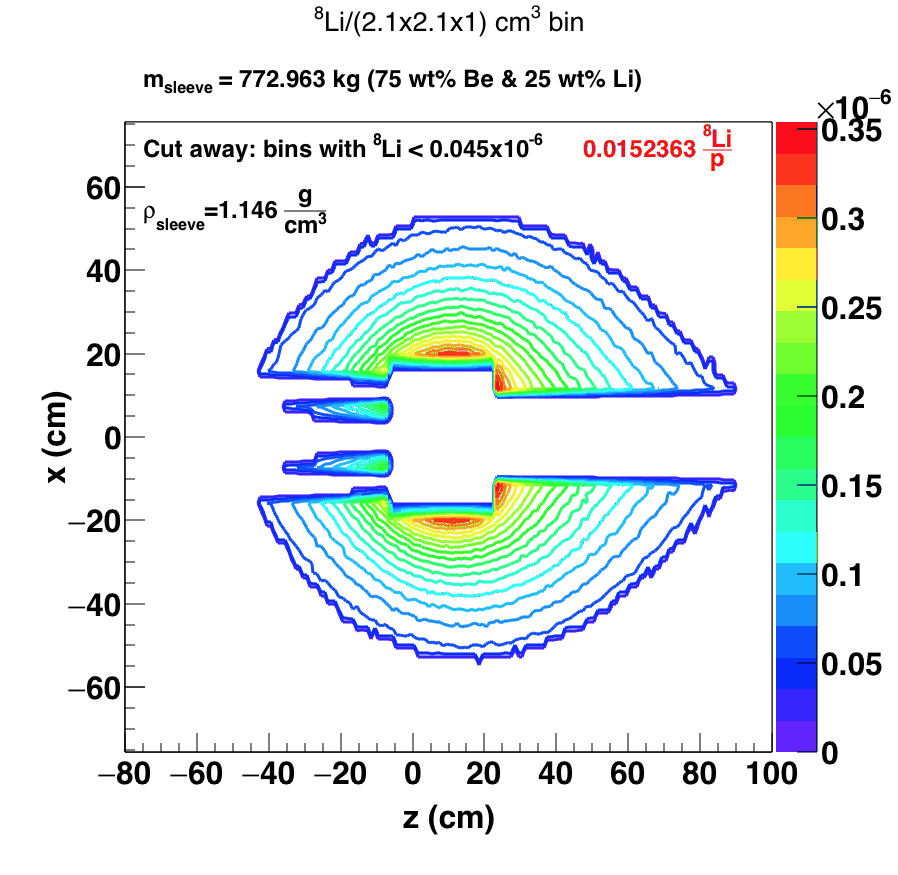}}\\
\hspace{1mm}
                     {\label{fig:f04}\includegraphics[width=59mm, height=67mm]{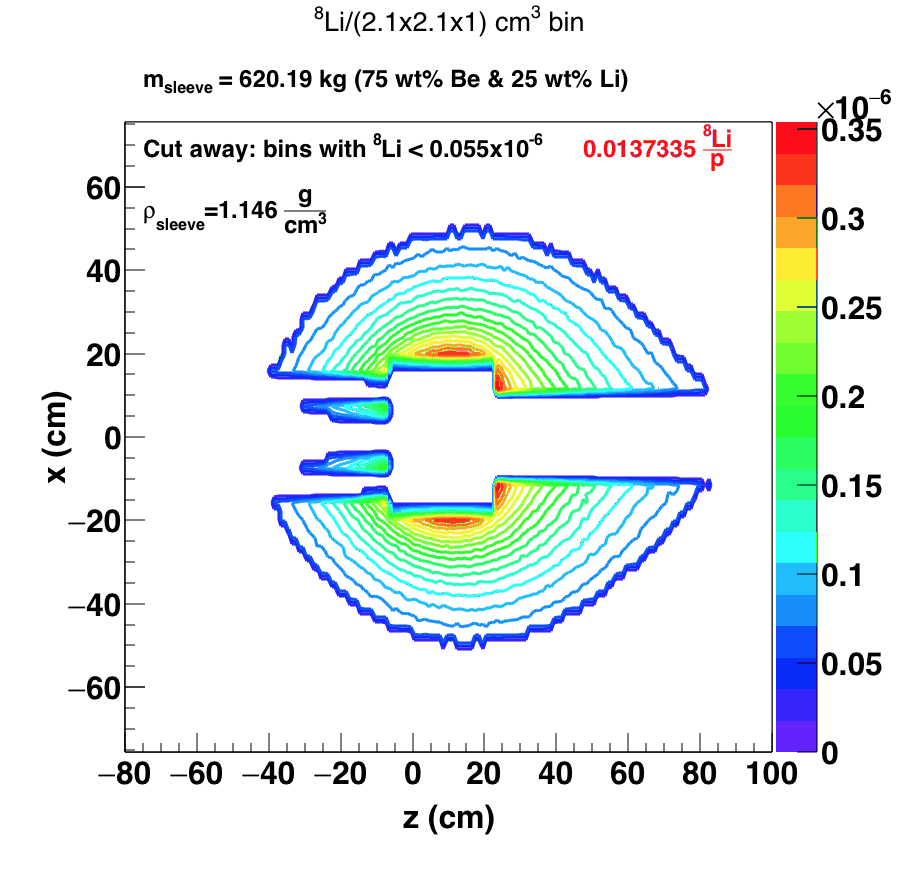}} 
                     {\label{fig:f05}\includegraphics[width=58mm, height=67mm]{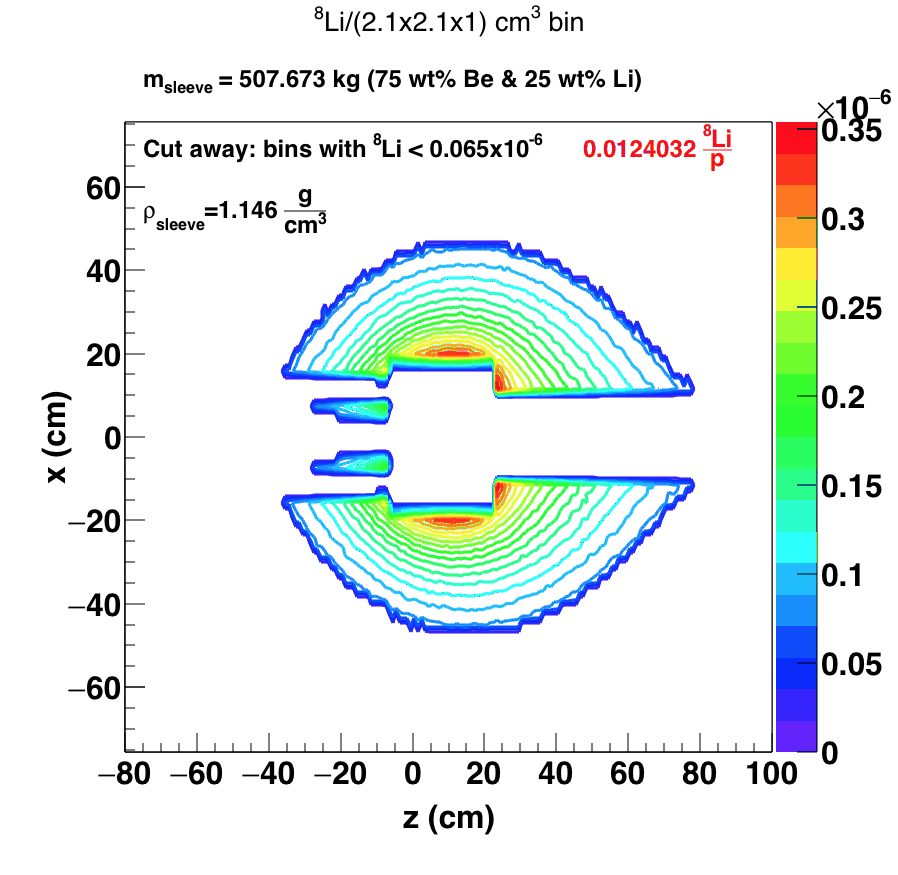}} 
                     {\label{fig:f06}\includegraphics[width=58mm, height=67mm]{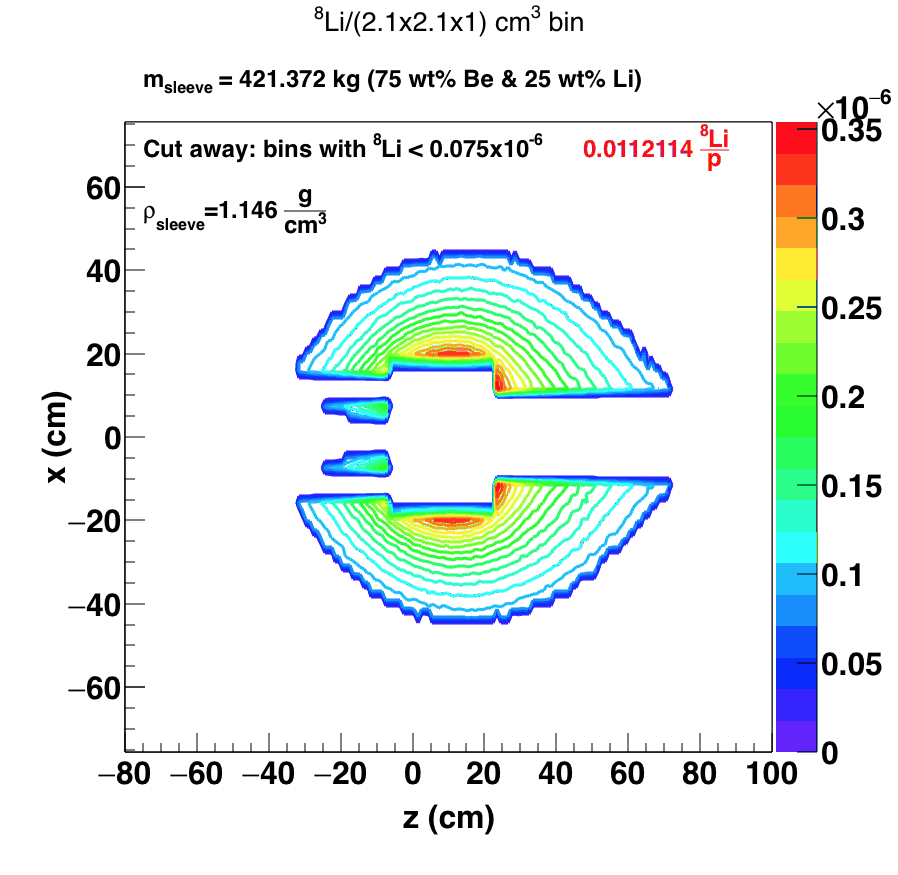}} \\
\hspace{1mm} 
                     {\label{fig:f07}\includegraphics[width=59mm, height=67mm]{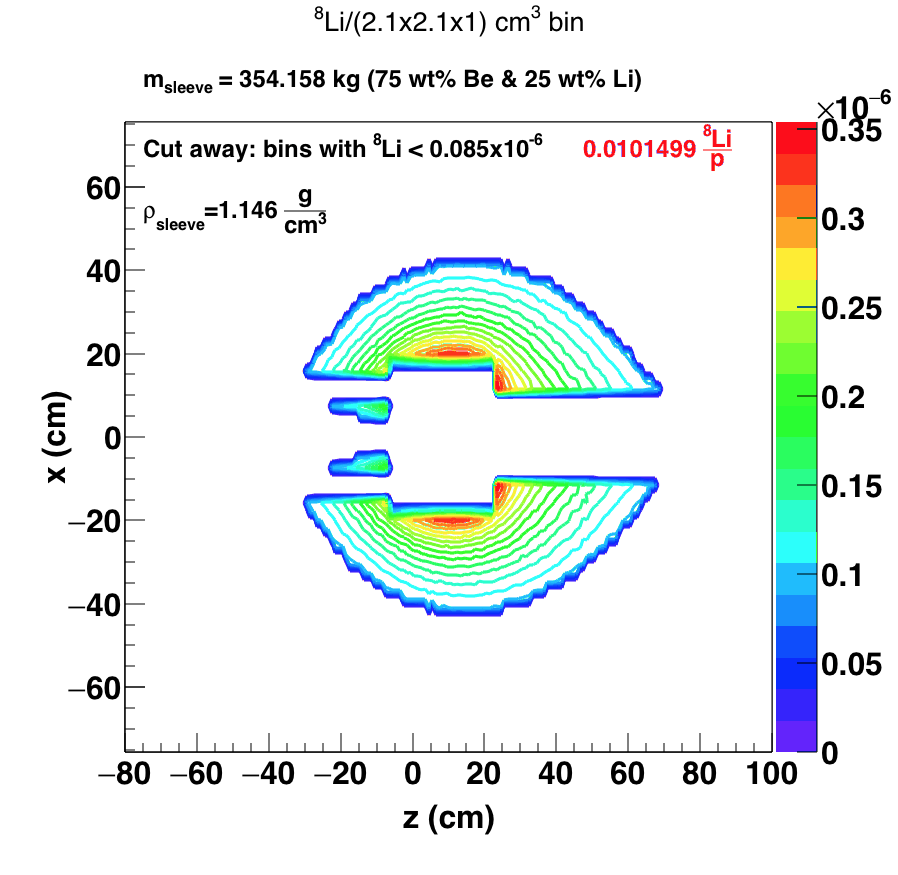}} 
                     {\label{fig:f08}\includegraphics[width=58mm, height=67mm]{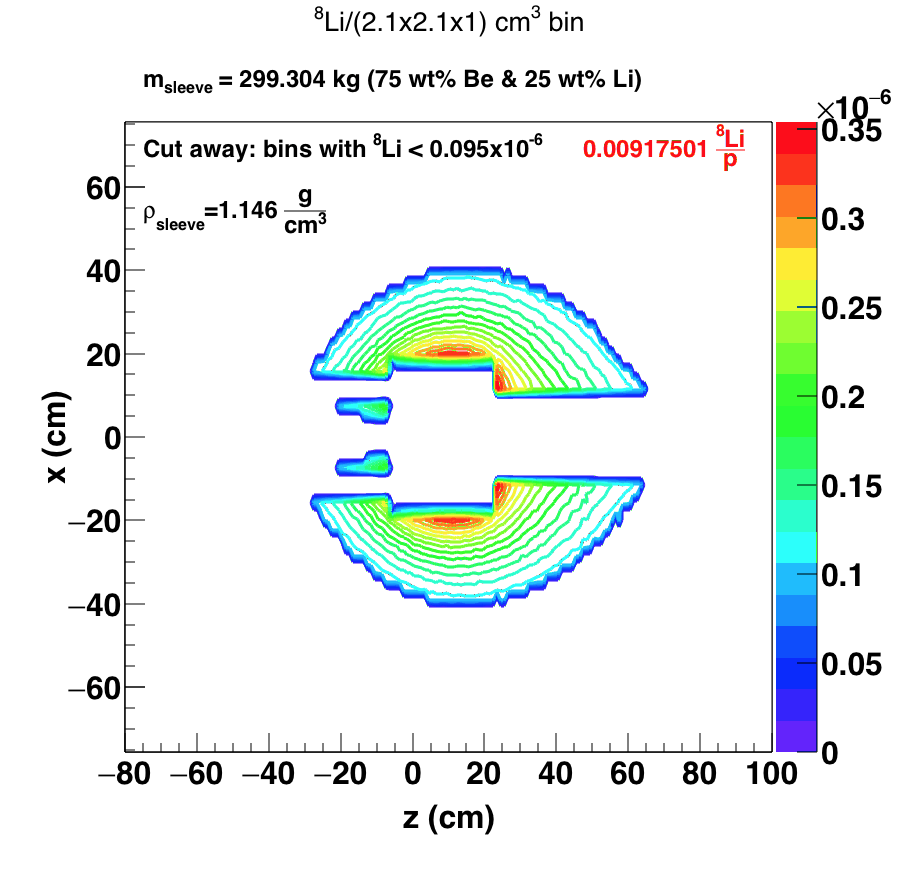}} 
                     {\label{fig:f09}\includegraphics[width=58mm, height=67mm]{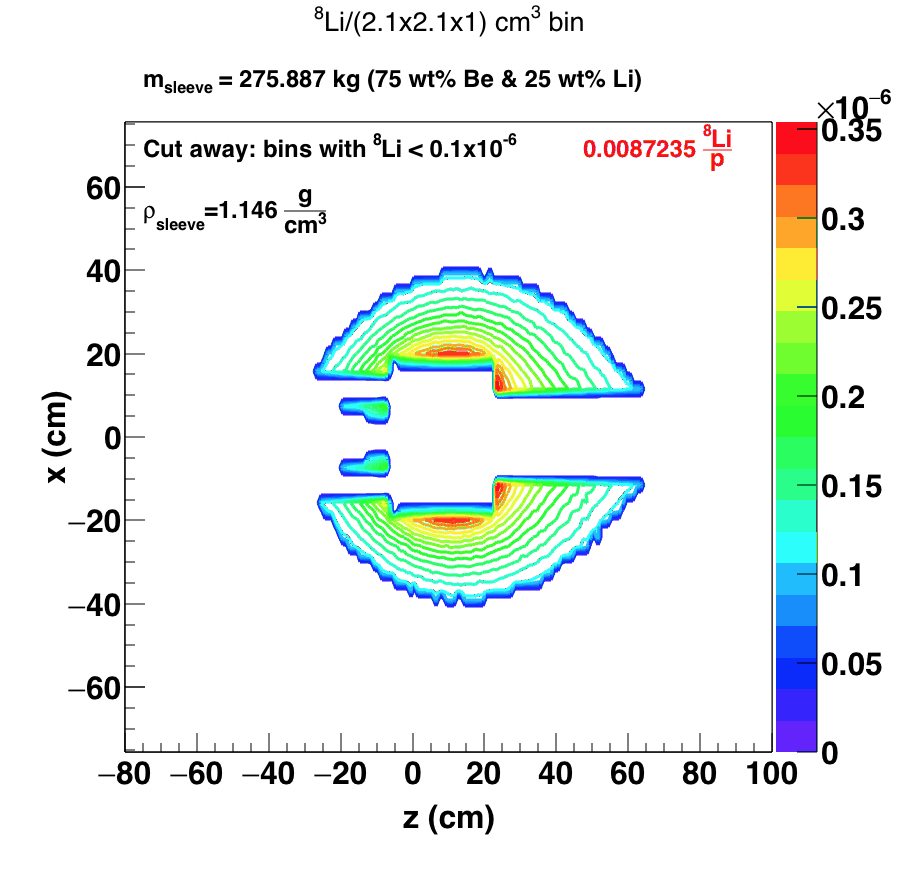}} \\
  \end{center}
  \caption{Optimization of the size and shape of the sleeve. The
    sleeve is made of a homogeneous mixture of Li and Be (Be mass
    fraction is 75$\%$). The parameters of interest are the mass of the sleeve and the
    corresponding $^{8}$Li yield. }
  \label{fig:f10}
\end{figure*}

With the optimum sleeve composition established in terms of maximizing the 
$^{8}$Li yield, the high cost of producing the enriched $^{7}$Li sleeve requires 
a second optimization study of the actual size and shape of the sleeve. 
Therefore, we simulated a large sleeve in order to determine
the regions inside the sleeve where the $^{8}$Li yield is below
certain threshold values. We then remove these regions while 
recording how this affects the total $^{8}$Li production. The sleeve radius 
was increased to 2~m, keeping the length at 1.9~m, and the material 
consists of 75$\%$ mass fraction Be and 25$\%$ mass fraction Li 
with${^7}$Li purity of 99.99$\%$. The density as well as the Be and 
Li mass fractions are maintained at constant values throughout this 
study. Figure~\ref{fig:f10} shows a section plane cut through the 
sleeve, showing the shape of the sleeve while the $^{8}$Li production 
yield threshold was varied. As the volume and mass of the sleeve 
were reduced, so was the overall $^{8}$Li production rate. This provides 
an evaluation of the overall cost of the sleeve versus total $^{8}$Li yield. 
Keeping the axes ranges fixed, one can see the continuous drop in the 
sleeve volume. The numerical values of the main parameters are also 
shown in Table~\ref{tab:yield_mass} for reference.

 \begin{table}[h!]
\centering
\caption{\label{tab:yield_mass} Correlation between the bin cut threshold 
  and the total mass of the sleeve, Li mass (kg), Yield $^{8}$Li/p
  for 99.99$\%$ $^{7}$Li purity, and scale to 99.995$\%$
  respectively.}
 %\begin{tabular}{>{\bfseries}cc>{\bfseries}cc}
\vspace{0.2cm}
%\begin{tabular}{ccccc}
\begin{tabular}{|c|l|l|l|l|l|}
\hline\hline
 \thead{Bin cut \\threshold}& \thead{ total mass (kg) }
  &  \thead {Li mass (kg)} &  \thead{ $^{8}$Li yield ($\%$)} 
   & \thead {Scale to\\ 99.995$\%$}\\   \hline
             0&18218.00&4554.50&2.679&3.531\\
3.00E-08&1148.00&287.00&1.796&2.367\\
3.50E-08&996.40&249.10&1.699&2.240\\
4.00E-08&874.00&218.50&1.608&2.120\\
4.50E-08&773.00&193.25&1.523&2.008\\
5.00E-08&689.50&172.38&1.445&1.905\\
5.50E-08&620.00&155.00&1.373&1.810\\
6.00E-08&560.00&140.00&1.305&1.720\\
6.50E-08&507.70&126.93&1.240&1.635\\
7.00E-08&461.70&115.43&1.179&1.554\\
7.50E-08&421.40&105.35&1.112&1.466\\
8.00E-08&385.50&96.38&1.066&1.405\\
8.50E-08&354.16&88.54&1.015&1.338\\
9.00E-08&325.20&81.30&0.965&1.272\\
9.50E-08&299.30&74.83&0.918&1.209\\
1.00E-07&275.90&68.98&0.872&1.149\\
\hline\hline
 \end{tabular}
 \end{table}

\noindent
These results show that a sleeve mass of 996~kg, corresponding
to a length of 130~cm and a radius of 60~cm, results in a
$^{8}$Li yield above 0.016~$^{8}$Li per incident proton. These results
also show the optimum shape and size required for the sleeve.

\begin{figure}[!ht] 
\centering
\includegraphics[width=0.46\textwidth]{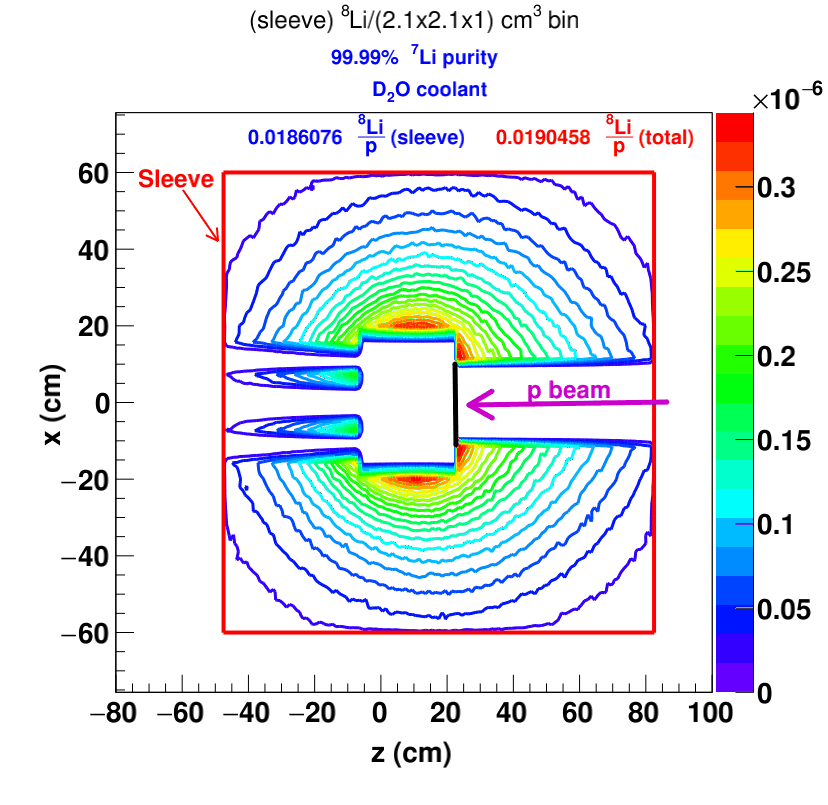}
\caption{The $^{8}$Li production with the new cylindrical sleeve.}
\label{fig:11}
\end{figure}

\noindent
Due to the technical difficulties in manufacturing a sleeve with a
teardrop shape as shown in Fig.~\ref{fig:f10}, a new, optimized cylindrical
shape using the dimensions determined above has been adopted. Also, the optimal sleeve
position with respect to the Be target has been determined. The results for this
cylindrical sleeve are shown in Fig.~\ref{fig:11}; the
$^{8}$Li yield is now 0.0186 $^{8}$Li per
incident proton, well above the initial $^{8}$Li yield which was
obtained for the previous FLiBe sleeve design. The total $^{8}$Li 
production in both the Be target and the sleeve is now 0.019~$^{8}$Li 
per incident proton. These results are for a 2~cm thick target. When
the thickness is changed to 1.7~cm, the total $^{8}$Li production becomes 
0.016~$^{8}$Li per incident proton, which meets the physics requirements for the IsoDAR experiment.

\section{Acknowledgements}

This work is supported by the National Science Foundation through
grants 1505858, 1707969 and 1707971.

\vspace{0.5 cm}

\section{Conclusion}

A very high flux of $\bar{\nu}_{e}$ will be produced by the decay of 
$^{8}$Li for the IsoDAR experiment. The focus of this paper has been on 
optimizing the target-sleeve design in terms of $^{8}$Li isotope production. 
As $^{8}$Li is produced overwhelmingly inside the sleeve, two materials are
considered for the sleeve that surrounds the Be target - an eutectic mixture of
lithium-fluoride and beryllium-floride (FLiBe) and a homogeneous mixture 
of Li-Be. A detailed study of neutron interactions on each chemical element 
inside the two sleeve materials shows that Li-Be performs better. The study 
shows that F in FLiBe, apart from drastically reducing the atomic number 
density of both Li and Be inside the sleeve, and therefore the number of 
neutron interactions with Li and Be, does not contribute to $^{8}$Li production 
and the number of secondary neutrons produced after each
neutron-F interaction is slightly decreased. The Be mass fraction for
which the $^{8}$Li yield is maximized is 75$\%$. This can be achieved
in practice by having a Li sleeve stacked with Be spheres with a
radius of up to 1 mm. With this sleeve material and Li-Be mass fractions 
as above, a size and shape sleeve optimization study was performed. By 
removing areas with $^{8}$Li yield below a certain threshold, a teardrop sleeve
shape was obtained. For a sleeve mass of of 996 kg, the $^{8}$Li yield
is above 0.016 $^{8}$Li per incident proton. With the new sleeve
design, the total $^{8}$Li yield is 0.019 $^{8}$Li per proton for a
2~cm thick target and 0.016 $^{8}$Li per proton for a
1.7~cm thick target.

%\nocite{*}

\end{document}